\documentclass{article}

\usepackage{arxiv}

\usepackage[utf8]{inputenc} 
\usepackage[T1]{fontenc}    
\usepackage{hyperref}       
\usepackage{url}            
\usepackage{booktabs}       
\usepackage{amsfonts}       
\usepackage{nicefrac}       
\usepackage{microtype}      
\usepackage{graphics}

\title{Semantic Segmentation of Thigh Muscle using 2.5D Deep Learning Network Trained with Limited Datasets}

\author{Hasnine Haquer\textsuperscript{1,2}\thanks{hasnine.haque@ge.com}, Masahiro Hashimoto\textsuperscript{2}, Nozomu Uetake\textsuperscript{1},Masahiro Jinzaki\textsuperscript{2} \\
		\textsuperscript{1}GE Healthcare, 4-7-127 Asahigaoka, Hino, 191-8503, Tokyo, Japan \\
		\textsuperscript{2}Dept. of Radiology,Keio University School of Medicine, 35 Shinanomachi, Shinjuku-ku, 160-8582, Tokyo, Japan
}

\date{}

\renewcommand{\headeright}{}
\renewcommand{\undertitle}{Technical Report}

\begin{document}
\maketitle

\begin{abstract}
\textbf{Purpose:} We propose a 2.5D deep learning neural network (DLNN) to automatically classify thigh muscle into 11 classes and evaluate its classification accuracy over 2D and 3D DLNN when trained with limited datasets. Enables operator invariant quantitative assessment of the thigh muscle volume change with respect to the disease progression.\newline
\textbf{Materials and methods:} Retrospective datasets consist of 48 thigh volume (TV) cropped from CT DICOM images. Cropped volumes were aligned with femur axis and resample into 2mm voxel-spacing. Proposed 2.5D U-Net consists of three 2D U-Net trained with axial, coronal and sagittal muscle slices respectively. A voting algorithm was used to combine the output of U-Nets to create final segmentation. 2.5D U-Net was trained on PC with 38 TV and the remaining 10 TV were used to evaluate segmentation accuracy of 10 classes within Thigh. The segmentation result of both left and right thigh were de-cropped to original CT volume space. Finally, segmentation accuracies were compared between proposed DLNN and 2D,3D U-Net.\newline
\textbf{Results:}Average segmentation DSC score accuracy of all classes with 2.5D U-Net as 91.18\% and Average Surface Distance (ASD) accuracy as $0.84\pm0.27mm$. We found, mean DSC score for 2D U-Net was 3.3\% lower than the that of 2.5D U-Net and mean DSC score of 3D U-Net was 5.7\% lower than that of 2.5D U-Net when trained with same datasets.\newline
\textbf{Conclusion:} We achieved a faster computationally efficient and automatic segmentation of thigh muscle into 11 classes with reasonable accuracy. Enables quantitative evaluation of muscle atrophy with disease progression.
\end{abstract}


\section{Introduction}
Thigh muscle composed of groups of muscle classes closely buddle together produces body heat using metabolic process. Segmentation and quantitative assessment of individual muscle classes in medical images provide valuable information for understanding their physiology, estimate overall risk of bone fracture  \cite{Kohout2013wrapping1,Takao2017fatty2} during rehabilitation. Semantic segmentation of the thigh muscles means to classifying each pixel that belongs to a specific substructure of muscle. Segmentation of muscles and their substructures ware used to calculate clinical parameters such as volume, as well as to define the search region for computer-aided detection tasks to improve their performance.

\begin{figure*}[h]
	\centering
	 \includegraphics{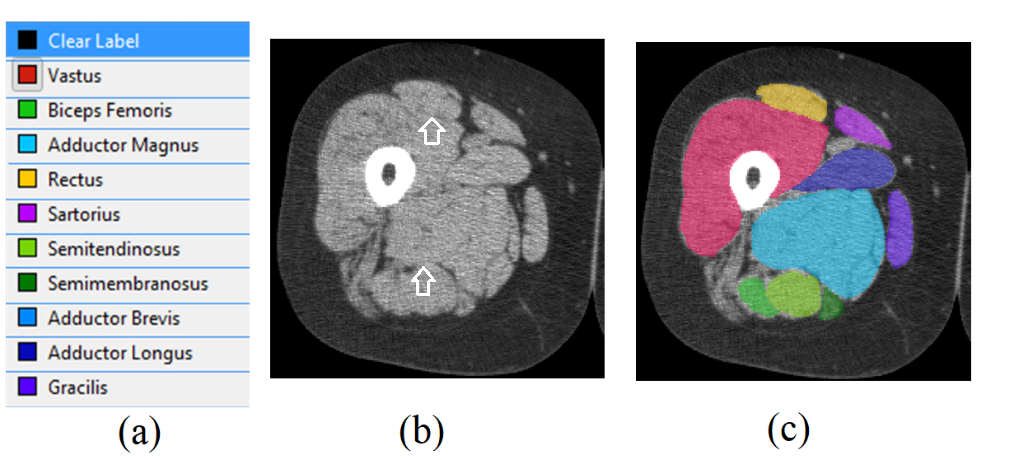}
	\caption{(a)Semantic muscle class names and their color coding. (b) typical CT axial slice of right thigh, white arrows indicate the not clear inter-muscle boundary. (c) Overlaid color-coded muscle segmentation on the CT image slice.}
	\label{fig:fig1}
\end{figure*}

\paragraph{} CT images ware most of the case available since patient undergoing orthopedic problems usually had CT scan as standard routine procedure \cite{Paul2005femoroacetabular3,Popuri2016skeletalmuscle4} and Manual delineation of multi class semantic labeling on thigh region consisting large number of thinner CT slices (1-2mm) is time consuming and suffers from inter- and intra-operator variability. Automatic muscle segmentation from CT images has number of challenges, like low soft tissue contrast and unclear inter-muscle boundary as shown in figure \ref{fig:fig1} (white arrows) and significant overlap between the Hounsfield unit (HU) value ranges of the muscle and surrounding tissues. A recent study on automated muscle segmentation from CT of the hip and thigh muscles using hierarchical multi-atlas method \cite{Yokota2018hierarchical5} into 11 and 9 classes respectively was reported and achieved good segmentation accuracy, however the computation time for segmentation (40 min) seems to be little long. Segmentation method based on deep neural network has demonstrated enormous success in improving diagnostic accuracy, speed of image interpretation, and clinical efficiency for a wide range of medical tasks, ranging from the interstitial pattern detection on chest CT \cite{Anthimopoulos2016Lung7} to bone age assesment on hand radiographs \cite{Lee2017assessment8}.

\paragraph{} Our contribution in this research was to solve the complex semantic muscle segmentation with a fully automated deep learning segmentation algorithm. Recently supervised learning (FCN), U-Net \cite{Ronneberger2015Unet9} became the state-of-the-art model for binary and semantic segmentation and widely used in two or three dimensions. Training a supervised model requires large number of ground truth (GT) semantic labeling data. Whenever training data is less in comparison to the number of parameters the model must learn, there is high chance of overfitting, because the model will learn too many rules for small dataset and might fail to generalize well to the unseen data. There has been always a debate whether to use 2DU-Net or 3DU-Net \cite{Ronneberger20163DUnet10} for volumetric segmentation.  Both 2D and 3D network models have their own advantages and disadvantages. For example, 2D network is simple less parameter and less efficient for 3D seg. Whereas 3D network has lots of trainable parameters and difficult to train. In this study a novel alternative model called 2.5D U-Net to solve almost all the challenges of semantic segmentation. Our objective is to segment thigh muscles in to 11 classes. Figure 1 shows the list of muscle classes and their respective color code for visualization. Background class consist of all the other tissues (bone, fat, vessels, etc.) other than defined 10 muscle classes. 

\section{Methods \& Materials}
\label{sec:methods}
\subsection{Segmentation Framework}
\begin{figure*}[h]
	\centering
		\includegraphics{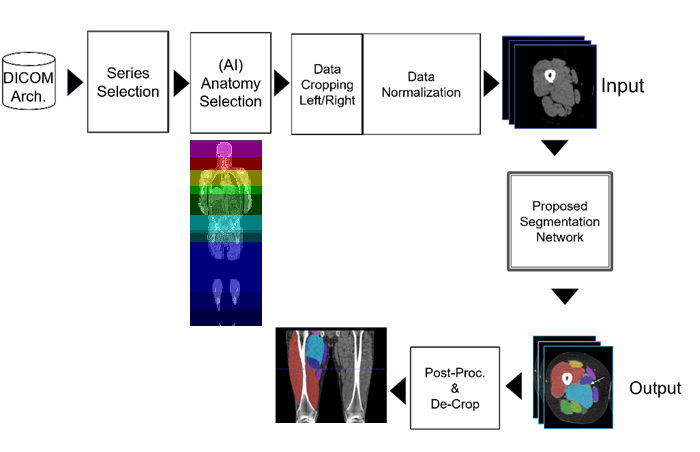}
	\caption{Overview of End to End Segmentation framework, starting form the raw DICOM input CT images and ending semantic segmentation results in input image space. An AI based anatomy selection classifier was used to extract only the slices ranges where the selected anatomy is present.}
	\label{fig:fig2}
\end{figure*}

For full automation of segmentation process, a generalized muscle segmentation framework was proposed which could be adapted to semantically segmentation of muscles in any other body parts. Figure \ref{fig:fig2} shows the overview of the framework. Input of the framework ware the raw DICOM images from a selected series. Input images were then processed by an AI based classifier \cite{Laszlo20163recognition11} to classify image slices(axial) into 12 organ class (None, Brain, Head-Neck, Chest, Abdomen-upper, Abdomen-lower, Pelvis-upper, Pelvis-center, Pelvis lower, Thigh, Shin, and Foot). Ranges of image slices which belong to the total thigh were automatically selected, cropped and normalized into left and right thigh and ready to be segment with the proposed neural network. The output of the segmentation results from the model was de-cropped and transformed back to the original input image space.

\subsection{Data Normalization}
Cropped Image slices which only belong to thigh region were processed morphologically to separate left and right thigh region from the CT images. In clinical environment, patient could be imaged while the legs are not in straight standard posture on CT gantry. To normalize the diverse leg orientation all the data were geometrically transformed (affine) in the orientation of anatomical long axis of femur bone. Input image dimension (height=128, width=128, slices=180) of dataset were kept same for the network model, a fixed FOV of 256x256x360 mm was cropped for each thigh volume and resamples into 2 $mm^{3}$ voxel spacing. Since muscles shapes and are mirrored symmetric between left and right thigh, left thigh image slices were mirrored left-right to normalize the muscle orientation as of right thigh. We employed a 100 HU to 250 HU intensity range normalization where muscle pixels were clearly visible.
\subsection{DLNN Model Definition and Loss Function} 
\begin{figure*}[h]
	\centering
		\includegraphics{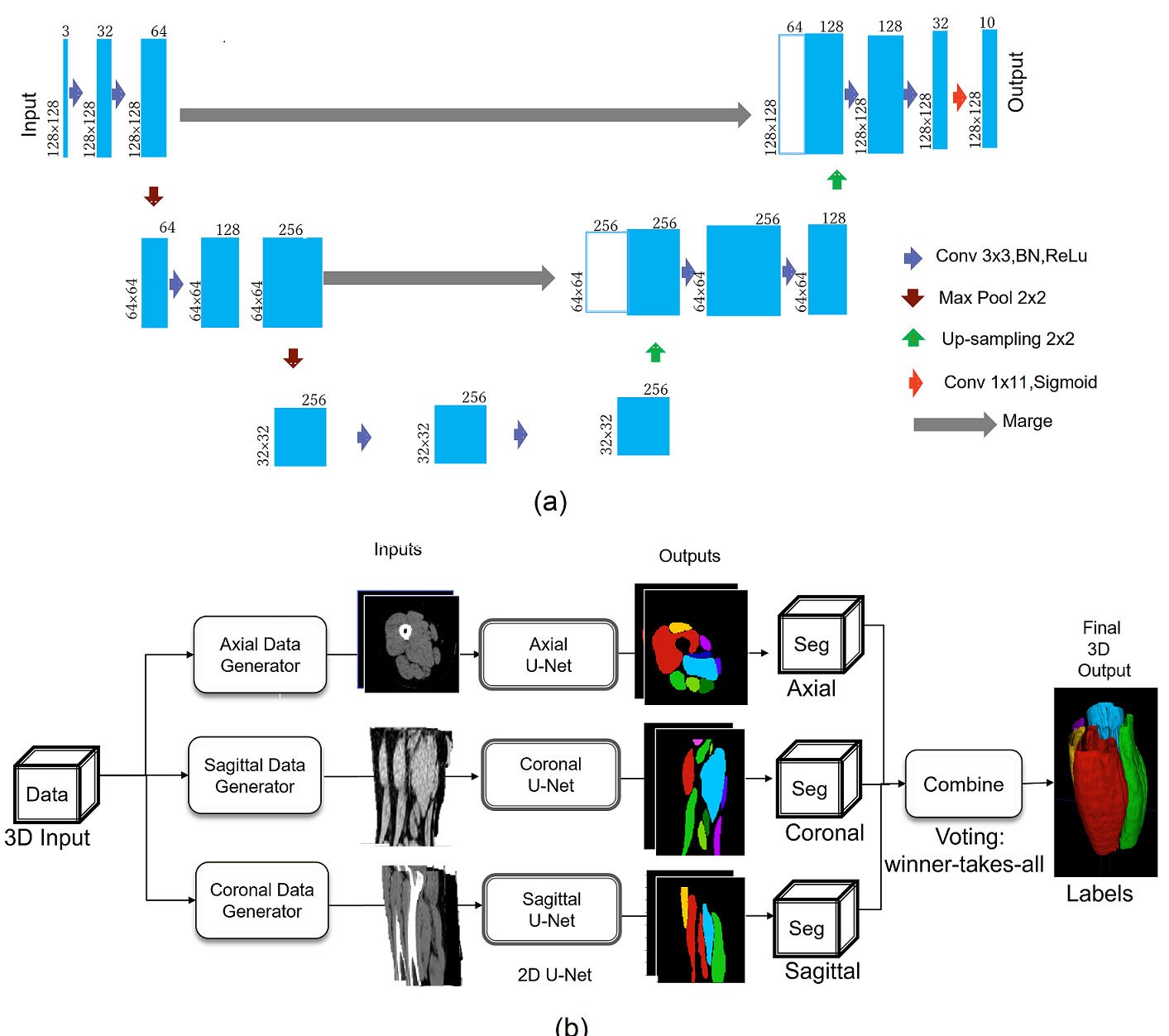}
	\caption{(a) Simplified 2D U-Net with only two skip connection. With the intension to keep the number of trainable parameters as low as possible (b) Overview of the proposed 2.5 U-Net model. 2D reformat generator generates 3 channels input for the Axial, Coronal and Sagittal U-Net model. Results of segmentation probability for all the slices from the three network models were combined with winners takes all voting rule to create the ensemble segmentation results.}
	\label{fig:fig3}
\end{figure*}

In this study we used the simplified 2D U-net architecture DLNN model shown in figure \ref{fig:fig3}(a) to keep the number of trainable parameters as less as possible with the intension to train them with smaller number of training datasets. Original U-Net model was proposed by Ronneberger et al. \cite{Ronneberger2015Unet9} as segmentation network, and widely used in medical image analysis. we proposed input of the segmentation model not only the slice to be segmented but also the its neighboring slices +- 4 mm away constituting three channels. The network model was trainined by minimizing area of predicted segmentation that do not overlap with the ground truth(actual) segmentation. Mathematically it can be represented as Dice similarity coefficient (DSC) value \cite{Dice1945Measures13} act as loss metric function of the model.

\subsection{2.5D U-Net Model} 
Our proposed model consists of three of these simplified 2D U-Nets and trained with images reformatted in three different orientation for segmentation. Figure \ref{fig:fig3}(b) shows the overview of the model. To segment total thigh volumes CT axial image slices were reformatted and generates series of 2D three channel axial, sagittal and coronal image slices and processed by respective simplified 2D U-Net network model.  Accumulating output segmentation probability map from all the image slices constitute the segmentation volume maps. In next step, segmentation probability volume maps from the three model were ensembled by winners takes all voting rule to create the final predicted semantic segmentation label. 2.5D U-Net model contains relatively much less number of parameter compare to 3D and can be trained with a smaller number of datasets to achieve greater segmentation accuracy and can be trained with more information on muscle’s sharp boundaries as compared to 2D U-Net.
\subsection{Train \& Test Model} 
In accordance with the Institutional Review Board at Keio University Hospital, 100 anonymized retrospective 64 row detector non-contrast retrospective CT image datasets were used for this study. Out of these data total of 48 thigh CT volumes from 24(Female:30, Male:8) patient with age range (17-87) years and median of 70 years were randomly chosen to train and test the model. The slice thickness of the CT datasets was between (1-2mm). We have excluded data with metal implants on the thigh regions, sever joint deformity and sever muscle atrophy (ALS patient). Ground truth annotation of the volumes were made by an expert radiologist (co-author). 10 CT volume out 48 thigh volumes were used to test the segmentation accuracy and remaining image data were augmented with $\pm10\%$ random scaling, $\pm10pix$  random shift and maximum $20\deg$ random rotation to train proposed the model in 100 epochs . Network parameters were trained and optimized with Adam ($\alpha=0.0001, \beta1=0.9, \beta2=0.999, lr=1e-4, decay=1e-3$) optimizer and average Dice similarity coefficient DSC \cite{Dice1945Measures13} as loss function between the predicted multi-class segmentation and the ground truth annotation. A post-processing of 3D connected component \cite{Samet1988Bintrees14} step was used to keep only the largest connected volumes to get final segmentation result.
\subsection{3D Evaluation}
The datasets used for evaluation of the proposed model were not used for training. Predicted muscle labels were quantitatively evaluated in 3D with respect to the ground truth labels of the test dataset in two ways. (1) total mismatch by amount of non-overlapping area between the predicted and ground truth label for all the slices as DSC score and (2) 3D shape mismatch of the predicted muscle surface points and ground truth muscle surface points as Average Similarity Distance (ASD) \cite{Styner2008segmentation16}. Finally, combined accuracy of the above two matrices was computed by taking average over all the muscle class. Additionally, we expressed the quantitative advantage of the proposed model over 2D U-Net model and 3D U-Net by the evaluation matric values when trained with same datasets. The matric values were plotted graphically by paired box and whiskers plot and statistical compaired by paired student t-test statistical significance score.

\section{Results}
\begin{figure*}[ht]
	\centering
		\includegraphics{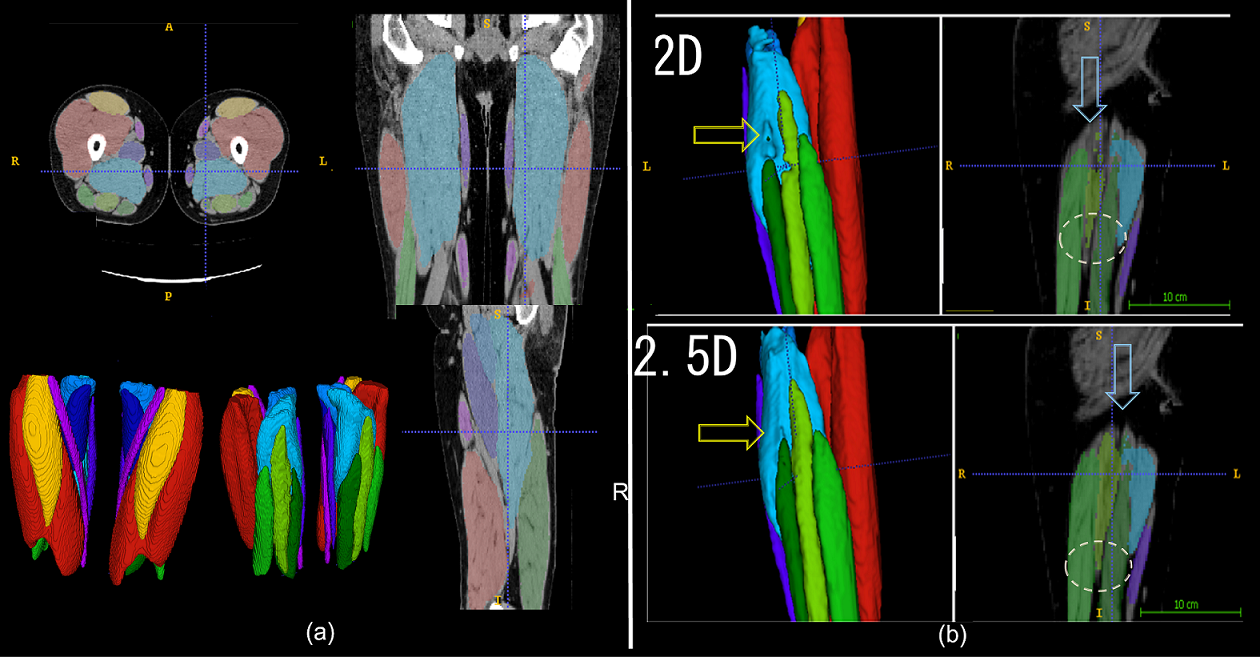}
	\caption{(a) Visual segmentation results 2.5D U-Net. Bottom left view shows surface of thigh muscles from two different view position, other three views shows the axial coronal and sagittal image plane overlaid with color coded muscle segmentation results. (b) Visible differences in segmentation results of 2D U-Net and 2.5D U-Net in 2D coronal plan view(left) and in 3D surface rendering view(right). Arrow head and circular region indicates the location where the clear segmentation result difference was observed.}
	\label{fig:fig4}
\end{figure*}

Segmentation evaluation was conducted over 10 test datasets, Figure \ref{fig:fig4}a shows visible the 11-class semantic segmentation result of thigh from one of the test dataset. Color coded classified muscles were overlaid on three orthogonal planes. On bottom corner of the figure shows the 3D surface rendering of the color-coded segmentation surface from two opposite viewpoints. On the figure 4b shows the visible comparison of segmentation results between 2D U-Net and the proposed 2.5D U-net. Left hand side view shows the surface rendering presentation and, on the right shows the 2D presentation of segmentation results. Region indicated by the arrow head and circle indicates the region where it is clearly indicates the improvement of segmentation results with the 2.5D U-Net.

\paragraph{}
Quantitative DSC score and ASD comparisons of the segmentation accuracy among 2D, 2.5D and 3D U-net model when train with same dataset are shown in figure 5. Paired box-and-whisker plots of the DSC accuracy score(top) and ASD value(bottom) for each class were created where * notation indicating statistical improvement in accuracy. On the right of the same figures shows the box-and-whisker plot of average DSC score comparison over all muscle classes. From this figure, it is observed that there exists a significant improvement of DSC score and ASD accuracy of the proposed 2.5D U-Net model over 2D U-Net model and reaches DSC score of 0.91. 2.5D U-Net Model shows superiority in DSC accuracy over 2D and 3D model. However, in 3D U-Net model DSC accuracy IQR is narrower than 2.5D U-Net model indicate its stable performance and may supersede the 2.5D U-Net model when trained with large number of datasets.  Also, it was observed from ASD plot that 2.5D U-Net model had significant improvement in distance-based accuracy compared to 2D U-Net model, however 3D U-net model showed lower mean ASD as compared to our proposed model.

\begin{figure*}[ht]
	\centering
		\includegraphics{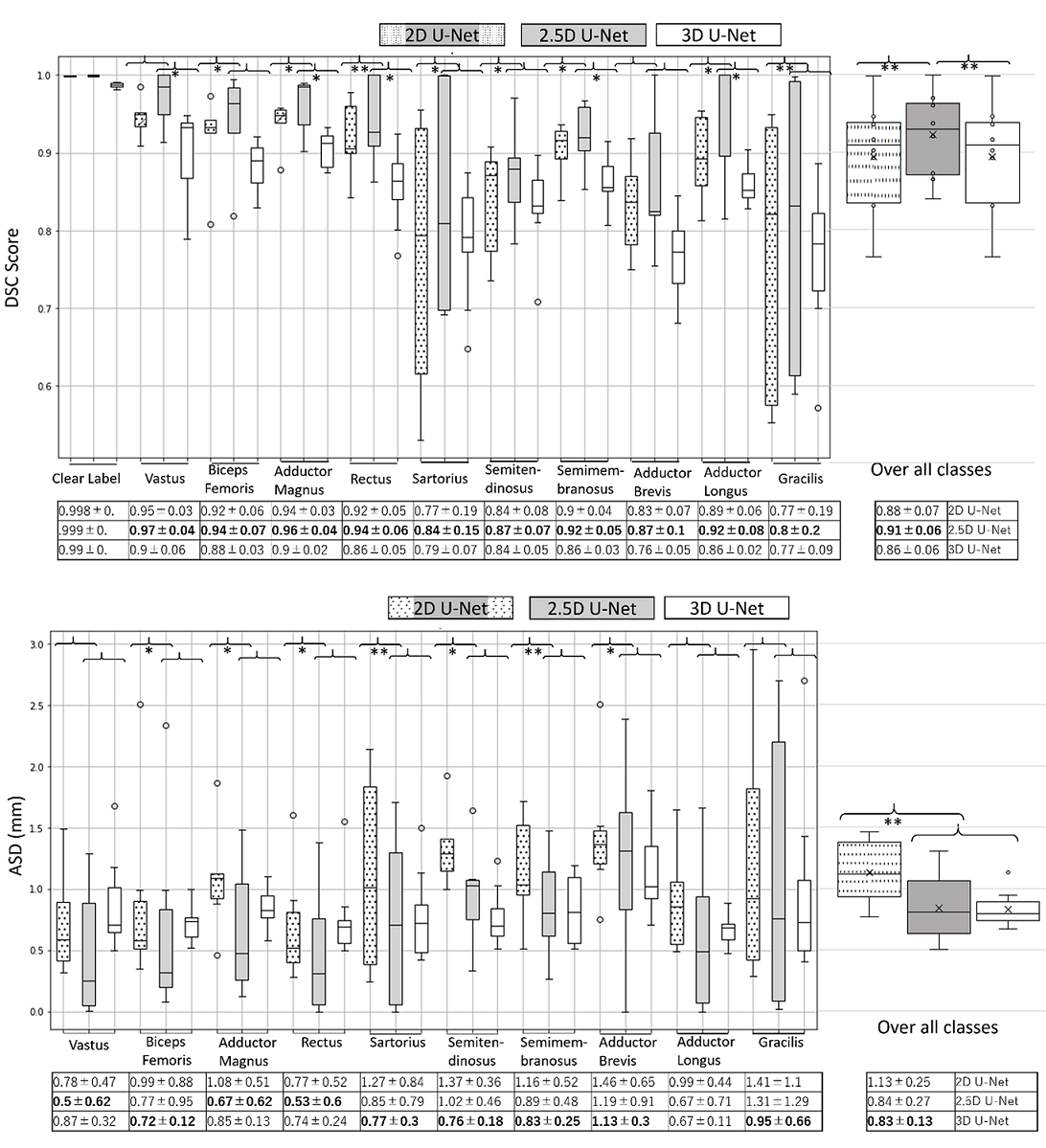}
	\caption{Box and whisker plot of DSC score comparison(top) and ASD comparison (bottom) between 2D U-Net, 2.5D U-Net and 3D U-Net and respective data table at the bottom. * indicates segmentation accuracy improvement between 2.5D U-Net and 2D U-Net with significance level less than 0.05 and ** indicate the significant accuracy improvement with significance level less than 0.01. Right side box plot shows the segmentation accuracy over all muscle classes.}
	\label{fig:fig5}
\end{figure*}

\section{Discussion \& Conclusion}
We have created an automated, deep learning system to automatically classify and segment the thigh muscle volume of clinical CT image slices in a reasonably short computation time of $1.08\pm0.007$ min per thigh volume on a GPU equipped Linux workstation (HP Xeon E5-2650 $v4 \times 2$, Memory 128GB, GPU: Tesla P100 $\times$ 4 OS: Ubuntu 16.04 LTS Python 3.6 Tensorflow 1.4 and Keras 2.0). This system achieves excellent overlap with ground truth segmented label images with an average of DSC score of $0.91\pm0.06$. Which is the best segmentation performance among 2D U-Net and 3D U-Net model when trained with same datasets. We demonstrated ASD segmentation accuracy of $0.84\pm0.27$ mm with the $1.61\pm0.70$ mm reported by Yokota et.al \cite{Yokota2018hierarchical5} using two-stage hierarchy method in 3D evaluation. Fully automated segmentation system can be embedded into the clinical environment to accelerate the quantification of muscle to provide advanced morphometric data on existing CT volumes. Which allows its adaptation of clinical research of automatic quantitative evaluation of muscle atrophy with the disease progression. Furthermore, we proposed an efficient solution for creating ground truth label images by manually editing segmentation results at the muscle boundary from initially trained U-Net network model with fewer training datasets. It helps to reduce manual workload of the create ground truth level of 48 thigh volumes. However, we found that segmentation DSC score were degraded for smaller muscle volume classes. In future, we need to improve the network model to handle variety of smaller muscle volumes and shrinking muscles cases due to muscle atrophy. Our proposed technique can be applied to segment muscle in other body parts or with different image modality.
\newpage

\bibliography{references}  
\bibliographystyle{IEEEtran}


\end{document}